\documentclass[aps,unsortedaddress,10pt,nobibnotes,notitlepage,twocolumn,longbibliography]{revtex4-1}

\usepackage[normalem]{ulem}

\usepackage[latin1]{inputenc}
\usepackage{graphicx}
\usepackage{float}
\usepackage{latexsym}
\usepackage{graphicx}
\usepackage{amssymb}
\usepackage{amsmath}
\usepackage{amsfonts}
\usepackage[colorlinks=true,citecolor=blue,hyperfootnotes=false]{hyperref}

\usepackage{cool}
\usepackage{dcolumn}
\usepackage{textcomp}
\usepackage[dvipsnames]{xcolor}
\usepackage{xfrac}
\usepackage{slashed}
\usepackage{multirow}
\usepackage{derivative}
\usepackage{tensor}
\usepackage{physics}
\usepackage{comment}

\def\figs/B{B}
\def\bea{\begin{eqnarray}}
\def\eea{\end{eqnarray}}
\def\bg{\begin{eqnarray}}
\def\nd{\end{eqnarray}}

\def\beq{\begin{equation}}
\def\eeq{\end{equation}}

\begin{document}

\title{Debye Screening of Non-Abelian Plasmas in Curved Spacetimes}

\author{Elba Alonso-Monsalve}
\email{elba\_am@mit.edu}

\author{David I.~Kaiser}
\email{dikaiser@mit.edu}

\affiliation{Department of Physics, Massachusetts Institute of Technology, Cambridge, MA 02139, USA}

\begin{abstract} Decades of analytic and computational work have demonstrated that a charge immersed in a hot plasma is screened. For both Abelian and non-Abelian interactions, the characteristic screening length $1/m_D$ is set by the so-called Debye mass $m_D \sim g_s T$, proportional to the plasma temperature $T$ and the dimensionless gauge coupling $g_s$. One of the most interesting naturally occurring examples is the quark-gluon plasma (QGP) that filled the early universe prior to the QCD confinement phase transition at $t_{\rm QCD} \sim 10^{-5}\,{\rm s}$. During this early epoch, regimes of strong spacetime curvature are of significant cosmological interest, such as near primordial black holes (PBHs). However, the typical description of Debye screening only applies within Minkowski spacetime, and is therefore insufficient to describe the dynamics of charged plasmas near PBHs or other primordial features. We construct an effective field theory for soft modes of the gauge field $A_\mu^a$ to give a full description of Debye screening in non-Abelian plasmas within arbitrary curved spacetimes, recovering a temperature-dependent Debye mass that exhibits gravitational redshift. We then apply our results to some scenarios of cosmological interest: an expanding FLRW universe and the vicinity of a PBH immersed in a hot QGP.
\end{abstract}

\date{\today}

\maketitle

\section{Introduction}

The screening of charges in hot plasmas, for both Abelian and non-Abelian interactions, has been a topic of significant interest for decades. Upon resumming hard thermal loops, the leading-order effects yield a characteristic screening length $\lambda_D (T) = 1 / m_D (T)$ set by the Debye mass $m_D (T) \sim g_s T$, where $g_s$ is the dimensionless gauge coupling strength and $T$ is the temperature of the plasma. (For reviews, see Refs.~\cite{Blaizot:2001nr,Litim:2001db,Kapusta:2006pm,Bazavov:2020teh}.)

The Debye mass $m_D (T)$ sets the characteristic scale for dynamics of collective excitations within the plasma. For example, collective oscillations of ``soft" modes, with momenta $k_{\rm soft} \sim m_D(T) \ll T$, propagate with frequency set by the plasma frequency $\omega_p \simeq m_D (T)$ \cite{Matinyan:1981dj,Klimov:1981ka,Klimov:1982bv,Weldon:1982aq,Braaten:1989kk,Braaten:1989mz,Braaten:1991gm,Blaizot:1993be,Blaizot:1993zk,Blaizot:1994da,Blaizot:1994vs,Blaizot:1995kg,Mrowczynski:2016etf}. Moreover, nontrivial spatial distributions of color charge among the soft gluons can form, with typical length-scale $1 / m_D (T)$ \cite{Manuel:2003zr,Manuel:2004gk}. These effects have been studied extensively in Minkowski spacetime, using both analytic techniques and lattice simulations \cite{Blaizot:2001nr,Litim:2001db,Kapusta:2006pm,Bazavov:2020teh}.

In this paper we consider Debye screening in hot plasmas within curved spacetimes, with particular interest in cosmological applications. For example, post-inflation reheating typically yields a universe filled with Standard Model particles in thermal equilibrium at a temperature as high as $T \sim {\cal O} (10^{14} \, {\rm GeV}) \sim {\cal O} (10^{27} \, K)$ by the time $t_{\rm therm} \sim {\cal O} (10^{-35} \, {\rm s})$ after the big bang \cite{Amin:2014eta,Allahverdi:2020bys}. Such temperatures are exponentially greater than the QCD confinement scale $\Lambda_{\rm QCD} \simeq 0.170 \, {\rm GeV}$. Hence at such early times, the universe was filled with a plasma of unconfined quarks and gluons, subject to the non-Abelian dynamics of QCD \cite{Mukhanov:2005sc,Boyanovsky:2006bf}.

As the universe expanded, the temperature of the plasma fell adiabatically: $T (t) a(t) \simeq {\rm constant}$, where $a(t)$ is the scale factor of the Friedmann-Lema\^{i}tre-Robertson-Walker (FLRW) line-element. During the radiation-dominated phase, $a (t) = (t / t_{\rm therm})^{1/2}$. The QCD confinement transition, after which quarks and gluons remained bound in color-neutral hadronic states, occurred at $t_{\rm QCD} \simeq 10^{-5} \, {\rm s}$, when $T (t_{\rm QCD}) \simeq \Lambda_{\rm QCD}$ \cite{Mukhanov:2005sc,Boyanovsky:2006bf}.

Within the range of times $t_{\rm therm} \ll t \ll t_{\rm QCD}$, additional phenomena of cosmological interest could have occurred. For example, if a significant population of primordial black holes formed at early times, with masses in the range $10^{17} \, {\rm g} \leq M (t_c) \leq 10^{22} \, {\rm g}$, they could account for the entire abundance of dark matter today \cite{Carr:2020xqk,Green:2020jor,Villanueva-Domingo:2021spv,Escriva:2022duf}. Given the dependence of the black holes' masses on the time of collapse $t_c$, the black-hole mass range of interest to account for the present dark-matter abundance corresponds to formation times $10^{-21} \, {\rm s} \leq t_c \leq 10^{-16} \, {\rm s}$. At these early times, the temperature of the plasma within which the primordial black holes formed would have been $10^5 \, {\rm GeV} \leq T (t_c) \leq 10^{7} \, {\rm GeV}$, for which $T (t_c) \gg T (t_{\rm QCD})$.

With such cosmological applications in mind, we study Debye screening in hot non-Abelian plasmas within arbitrary curved spacetimes, including spacetimes that need not be homogeneous and isotropic (such as in the vicinity of a primordial black hole). In such general spacetimes, the effective temperature of the plasma can develop spatial gradients, $T \rightarrow T (x)$, akin to the familiar Tolman temperature \cite{Tolman:1930zza,Tolman:1930ona,Santiago:2018lcy}. We identify corrections to the induced current $j_\mu^a (x)$ and to the Debye mass $m_D (T (x))$ from spacetime curvature. Much as in Minkowski spacetime, the components $A^a_0 (x)$ of the gauge field acquire an effective mass in the plasma proportional to $m_D (T (x))$, whereas components $A^a_i (x)$ remain massless. (See, e.g., Refs.~\cite{Holcomb,Dettmann:1993zz,Prokopec:2003bx,Prokopec:2003tm,Dodin:2010zz,Popov:2017xut,Cannizzaro:2020uap,Cannizzaro:2021zbp,Feng:2022evy} on the behavior of Abelian plasmas within curved spacetimes.)

In Section \ref{sec:EFT}, we introduce an effective theory for soft modes within the plasma, generalizing the elegant formalism reviewed in Refs.~\cite{Blaizot:2001nr,Litim:2001db} for application to curved spacetimes. In Section \ref{sec:current} we evaluate the induced current $j_\mu^a (x)$ for the soft modes that arises from interactions with high-energy quarks and gluons in the plasma. Section \ref{sec:DebyeMass} considers corrections to the usual expression for the Debye mass $m_D (T)$ arising from spacetime curvature. In Section \ref{sec:applications} we apply these expressions for $j_\mu^a (x)$ and $m_D (T(x))$ to some scenarios of cosmological interest.

We restrict attention to $(3+1)$ spacetime dimensions and adopt the metric signature $(-,+,+,+)$. Greek letters $\mu, \nu = 0, 1, 2, 3$ label spacetime indices, while Latin letters $i, j = 1, 2, 3$ label spatial indices. The color-charge indices associated with the adjoint representation of the gauge group ${\rm SU} (N_c)$ range over $a, b, c = 1, 2, ... , N_c^2 - 1$ and are raised and lowered with $\delta_{ab}$. The generators $T^a$ of the Lie algebra for ${\rm SU} (N_c)$ satisfy $[ T^a, T^b ] = i f^{abc} T^c$, where $f^{abc}$ are the totally antisymmetric structure constants, and the generators are normalized such that $2 \, {\rm Tr} (T^a T^b) = \delta^{ab}$. We adopt natural units in which $c = \hbar = k_B = 1$, in terms of which the reduced Planck mass is given by $M_{\rm pl} \equiv 1/ \sqrt{ 8 \pi G} = 2.43 \times 10^{18}$ GeV.

\section{Effective Theory for Soft Modes}
\label{sec:EFT}

At early times $t \ll t_{\rm QCD}$, when the temperature of the plasma filling the universe satisfied $T \gg \Lambda_{\rm QCD}$, the number density of gluon degrees of freedom likely dominated those of quarks: gluons can radiate gluons at tree level, and bosonic statistics allow gluon number densities per mode to scale as $n_k \sim 1 / \alpha_s$, where $\alpha_s \equiv g_s^2 / (4 \pi)$ and $g_s$ is the QCD coupling constant \cite{Gyulassy:2004zy}. Given the running of $\alpha_s$ to lower values at higher energies, the number density of gluons could have greatly exceeded the number density of quarks within the plasma at early times \cite{Blaizot:2001nr,Bazavov:2020teh}. Similar behavior has been observed in the plasmas that form immediately after relativistic heavy-ion collisions \cite{Scardina:2012mik,Gyulassy:2004zy}. Hence we expect gluons to dominate the dynamics, and consider the effective action for a Yang-Mills model with gauge group ${\rm SU} (N_c)$ and field strength tensor
\beq
F_{\mu\nu}^a = \nabla_\mu A_\nu^a - \nabla_\nu A_\mu^a + g_s f^{abc} A_\mu^b A_\nu^c ,
\label{Fmndef}
\eeq
where $\nabla_\mu$ is the (spacetime) covariant derivative associated with metric $g_{\mu\nu} (x)$.

Following the approach reviewed in Refs.~\cite{Blaizot:2001nr,Litim:2001db}, we construct an effective theory for long-wavelength excitations in the plasma within the high-temperature limit. In particular, we consider the effective action for soft modes $\tilde{A}_\mu^a$ with typical momenta $k_{\rm soft} \sim g_s T$. Thermal and quantum corrections to the behavior of the soft modes are dominated by quanta $a_\mu^a$ with momenta $k_{\rm hard} \sim T$. We therefore write $A_\mu^a = \tilde{A}_\mu^a + a_{\mu}^a$, integrate out the high-momentum modes $a_\mu^a$, and drop the tilde on the soft modes $\tilde{A}_\mu^a \rightarrow A_\mu^a$.

As described in Refs.~\cite{Blaizot:2001nr,Litim:2001db}, the soft modes $A_\mu^a$ have large occupation numbers per mode and hence behave as effectively classical fields. The behavior of the soft modes therefore yields the mean-field dynamics for the system on length-scales $\lambda \geq 1 / k_{\rm soft} \sim 1 / (g_s T)$. We follow the background-field method of Ref.~\cite{Blaizot:2001nr}, whereby we choose to decompose the action of the gauge symmetry on $a_\mu^a$ and $A_\mu^a$ in such a way that leaves the soft modes $A_\mu^a$ unchanged. One example is to use a generalized Coulomb-type gauge-fixing term in the full theory, $\nabla_i a_a^i - g_s f^{abc} A_i^b a_c^i$ \cite{Blaizot:2001nr}. No additional gauge-fixing terms or ghosts are then required in the effective action for the soft modes.

For long length-scales $\lambda \geq 1 / k_{\rm soft} \gg 1 / k_{\rm hard} \sim 1 / T$, the dynamics may be described by the effective action
\beq
S_{\rm eff} = \int d^4 x \sqrt{-g} \left[ \frac{ M_{\rm pl}^2}{2} R - \frac{1}{4} F_{\mu\nu}^a F_a^{\mu\nu} - j_\mu^a A_a^\mu  + {\cal L}_{\rm fluid}\right] ,
\label{Seff}
\eeq
where $j_\mu^a (x)$ is the induced current generated by (non-Abelian) self-interactions with the high-momentum modes $a_\mu^a (x)$; the induced current $j_\mu^a$ also includes contributrions from high-momentum quarks. The term ${\cal L}_{\rm fluid}$ represents the contributions to the evolution of the spacetime curvature from constituents other than the soft modes. For example, the high-momentum modes in the plasma (coarse-grained over a length-scale $\lambda \gg 1 / k_{\rm hard}$) behave as a neutral fluid with a radiation-dominated equation of state, and thereby contribute to the time evolution of the scale factor $a(t)$ in an FLRW background, while masses $M$ of compact objects, such as black holes, influence the spacetime curvature in their vicinity. Each of these contributions affects $g_{\mu\nu} (x)$ and hence impacts the dynamics of the soft modes $A_\mu^a (x)$.

Varying $S_{\rm eff}$ with respect to $A_\mu^a$ yields the equations of motion
\beq
\begin{split}
D^\mu F^a_{\mu\nu}  = j_{\nu}^a ,
\end{split}
\label{Aeom}
\eeq
where the covariant derivative $D_\mu$ acting on a spacetime tensor that transforms in the adjoint representation of ${\rm SU} (N_c)$, $X^a_{\nu_1 \cdots \nu_n}$, is defined as
\beq
D_\mu X^a_{\nu_1 \cdots \nu_n} \equiv \nabla_\mu X^a_{\nu_1 \cdots \nu_n} + g_s f^{abc} A^b_\mu X^c_{\nu_1 \cdots \nu_n}.
\label{covderivdef}
\eeq
The left-hand side of Eq.~(\ref{Aeom}) can be expanded as
\beq
\begin{split}
D^\mu F^a_{\mu\nu} &= \nabla^\mu F^a_{\mu\nu} + g_s f^{abc} A^\mu_b F^c_{\mu\nu} \\
&=D^{\mu}D_{\mu}A_{\nu}^a - g_s f^{abc}A^{\mu}_b\nabla_{\nu}A_{\mu}^c -R_{\mu\nu}A^{\mu}_a ,
\end{split}
\label{Aeom2}
\eeq
where $R_{\mu\nu}$ is the Ricci tensor for the background spacetime. The equations of motion imply that the induced current $j_\mu^a$ must be covariantly conserved with respect to both the spacetime curvature and the ${\rm SU} (N_c)$ gauge group:
\beq
D^\mu  j_\mu^a (x) = 0 ,
\label{jcons}
\eeq
for each color $a$.

The contribution to the energy-momentum tensor from the Yang-Mills field is given, as usual, by
\beq
T^{(A)}_{\mu\nu} = F^a_{\mu \lambda} F_\nu^{a \> \lambda} - \frac{1}{4} g_{\mu\nu} F_a^{\lambda\sigma} F^a_{\lambda \sigma} ,
\label{TYMgeneral}
\eeq
while the additional fluid contributes 
\beq
T^{\rm fluid}_{\mu\nu} = - \frac{2}{\sqrt{-g} } \fdv{\cal L_{\rm fluid}}{g^{\mu\nu}} .
\label{Tmnfluid}
\eeq
These terms satisfy the usual covariant conservation equations 
\beq
\begin{split}
\nabla_\mu T^{\mu\nu}_{(A)} &= j_\mu^a F^{\mu\nu}_a  , \\
\nabla_\mu T^{\mu\nu}_{\rm fluid} &= 0
\end{split}
\label{nablaTmn}
\eeq
in the presence of the induced current $j_\mu^a$.

\section{Induced Current}
\label{sec:current}

The induced current $j_\mu^a (x)$ for the soft modes $A_\mu^a (x)$ arises from interactions with high-momentum quarks and gluons in the plasma. Lattice simulations have confirmed the analytic expectation that for temperatures $T \gg T_{\rm QCD}$, high-energy quarks and gluons attain an equilibrium equation of state akin to that of a gas of non-interacting, massless particles. In particular, the so-called ``trace anomaly," $(\rho - 3P) / T^4$, where $\rho$ is the energy density and $P$ the pressure of particles in the plasma, tends rapidly toward zero for $T \gg T_{\rm QCD}$, and therefore $\rho\approx 3P$. (See, e.g., Refs.~\cite{Borsanyi:2012ve,Borsanyi:2013bia,HotQCD:2014kol,Pasechnik:2016wkt}.)

At high temperatures, the soft quantum fields of the plasma are well approximated by classical fields, due to their large occupation numbers per mode. Furthermore, the hard (high-momentum) modes can be approximated by an ensemble of classical point particles to leading order in the coupling $g_s$, since they are weakly interacting. Specifically, the effects of the high-momentum particles on the soft modes within the plasma are dominated by processes involving ``hard thermal loops": one-loop diagrams with arbitrary numbers of low-momentum external legs (with $k_{\rm soft} \sim g_s T$) and hard internal momenta (with $k_{\rm hard} \sim T$) \cite{Pisarski:1988vd,Braaten:1989mz,Braaten:1990az,Braaten:1990it,Braaten:1991gm}. These effects can be analyzed within a mean-field approximation of the (truncated) Schwinger-Dyson equations \cite{Blaizot:1993be,Blaizot:1993zk,Blaizot:1994vs,Blaizot:1995kg,Blaizot:2001nr,Mrowczynski:2016etf}, or by simply adopting classical transport equations for high-momentum particles \cite{Litim:2001db,Kelly:1994ig,Kelly:1994dh,Pisarski:1997cp,Mrowczynski:2016etf}. In this section we adapt the latter approach to arbitrary curved spacetimes. (See also Ref.~\cite{Brandt:1994mv}.)

The equations of motion for a classical point particle with color charge in a soft gauge-field background are well known
\cite{Wong:1970fu,Linden:1995bt,Kelly:1994dh,Brandt:1994mv}:
\beq
\begin{split}
    \dv{x^\mu}{\lambda} &= P^\mu , \\
    \dv{P_\mu}{\lambda} &= - \frac{1}{2} \pdv{g^{\alpha \beta}}{x^\mu} P_\alpha P_\beta - g_s Q^a F_{\mu\nu}^a P^\nu , \\ 
    \dv{Q^a}{\lambda} &= - g_s f^{abc} A_\mu^b Q^c P^\mu  .
\end{split}
\label{wongeq}
\eeq
Here $x^\mu$ is the position of the particle, $P^\mu = \dv*{x^\mu}{\lambda}$ its kinetic 4-momentum, $\lambda$ an affine parameter, and $Q^a$ its ${\rm SU}(N_c)$ charge. We take advantage of the freedom to linearly rescale the affine parameter $\lambda$ so that $\dv*{x^\mu}{\lambda}$ has units of energy. (We do not explicitly rescale by the particle mass $m$ so that our formalism can be applied to massless particles.) Notice that, due to non-Abelian self-interactions, the charges $Q^a$ are dynamical, unlike in electromagnetism.

The induced current $j^\mu_a (x)$, summed over all hard particles, is given by \cite{Blaizot:2001nr,Litim:2001db}
\beq
j^\mu_a (x) = g_s \int dQ \, d\omega \, \dv{x^\mu}{\lambda} Q_a \delta f (x, p, Q) ,
\label{jdef}
\eeq
where $\delta f (x, p, Q)$ represents the deviation from equilibrium of the distribution function for the high-momentum charge-carrying particles. The integration measure includes the momentum volume form $d\omega$ and the measure for the space of color charges $dQ$, subject to the physical constraints of on-shell mass condition, positivity of energy, and conservation of the $N_c-1$ group Casimirs. The momentum measure is given below, in Eq.~(\ref{domega1}). The phase-space structure for the color-charge degrees of freedom is unaffected by the spacetime structure, so we may use the now-standard parameterization developed for previous studies within Minkowski spacetime, as reviewed, e.g., in Ref.~\cite{Litim:2001db}; an explicit construction for the cases SU(2) and SU(3) is available in Section III of Ref.~\cite{Kelly:1994dh}. In particular, for the gauge group SU(2) the color-space measure is
\beq
dQ = d^3Q\,c_R\,\delta(Q^aQ^a-q_2) ,
\eeq
and for the gauge group SU(3) we have
\beq
dQ = d^8Q\,c_R\,\delta(Q^aQ^a-q_2)\delta(d_{abc}Q^aQ^bQ^c-q_3) .
\label{dQ}
\eeq
For the SU(3) case, the $d^{abc}$ are the totally symmetric constants given by $d^{abc} = 2 \, {\rm Tr} ( \{ T^a, T^b \}, T^c)$. The representation-dependent constant $c_R$ is fixed by the normalization $\int dQ=1$, while the constants $q_2,q_3$ are further fixed by the first and second Casimirs, respectively. Thus, the integration over color charges serves to enforce the conservation of the Casimir invariants \cite{Litim:2001db, Kelly:1994dh,Kelly:1994ig}.

We consider small departures from equilibrium, and therefore work perturbatively in powers of the coupling $g_s$, so we may write the full distribution function as
\beq
f (x, p, Q) = f^{(0)} (x,p) + g_s f^{(1)} (x, p, Q) + {\cal O} (g_s^2) .
\label{fseries}
\eeq
The relevant dynamics for the induced current $j^\mu_a (x)$ in Eq.~(\ref{jdef}) are captured to leading order by identifying $\delta f (x, p, Q) = g_s f^{(1)} (x,p, Q)$, as demonstrated by Refs.~\cite{Kelly:1994dh,Kelly:1994ig}.
The dynamical evolution of $\delta f(x,p,Q)$ is governed by the collisionless Boltzmann equation for the full distribution function $f(x,p,Q)$, which in the framework of Hamiltonian mechanics corresponds to the conservation of $f(x,p,Q)$ along dynamical trajectories (Hamiltonian flow) in phase space. 

To solve the Boltzmann equation, it will be convenient to have an explicit Hamiltonian formulation of the dynamics of the high-momentum particles. To this end, we first construct an action which yields the correct equations of motion, Eq.~(\ref{wongeq}). This is nontrivial because there is no such action involving the charges $Q^a$ directly as dynamical variables. To bypass this problem, 
following Refs.~\cite{Balachandran:1976ya,Barducci:1976xq,Linden:1995bt}, we introduce new dynamical variables $q^a$ which, like $Q^a$, transform under the adjoint representation of SU($N_c$), but which are anticommuting (Grassmann-valued). These new variables $q^a$ are useful as an intermediate step, in terms of which we may construct the charges $Q^a$ as
\beq
Q^a = -\frac{i}{2} f^{abc} q^b q^c .
\label{Qqdef}
\eeq
An action for a single high-momentum particle that yields the appropriate equations of motion may then be written in terms of the dynamical variables $x^\mu$ and $q^a$ as
\beq
\begin{split}
S_{\rm 1p} = \int d\lambda &\bigg\{ \frac{1}{2} g_{\mu\nu} \dv{x^\mu}{\lambda} \dv{x^\nu}{\lambda} +\frac{i}{2} q^a \dv{q^a}{\lambda} \\
&\quad  - \frac{i}{2} g_s f^{abc} A_\mu^a q^b q^c \dv{x^\mu}{\lambda} \bigg\} .
\end{split}
\label{Shm}
\eeq
The corresponding Hamiltonian is simply
\beq
H = \frac{1}{2} g^{\mu\nu} P_\mu P_\nu .
\label{Ham}
\eeq
The kinetic momentum $P^\mu \equiv \dv*{x^\mu}{\lambda}$ is related to the canonical momentum $p_\mu$, conjugate to $x^\mu$, by
\beq
P_\mu  = p_\mu - g_s Q^a A_\mu^a . 
\label{pcan}
\eeq
Then Hamilton's equations are Eq.~(\ref{wongeq}), as desired. Dynamical evolution in phase space is generated by the Liouville vector field $X_H$:
\beq
\begin{split}
X_H \equiv \dv{}{\lambda} &= \dv{x^\mu}{\lambda} \pdv{}{x^\mu} + \dv{p_\mu}{\lambda} \pdv{}{p_\mu} + \dv{Q^a}{\lambda}\pdv{}{Q^a} \\
&= P^\mu \pdv{}{x^\mu}\bigg\vert_{P} - \frac{1}{2} \pdv{g^{\alpha \beta}}{x^\mu} \bigg\vert_{P} P_\alpha P_\beta \pdv{}{P_\mu}\\
&\quad\quad - g_s Q^a P^\mu F_{\mu\nu}^a \pdv{}{P_\nu} \\
&\quad\quad  - g_s f^{abc} A^b_\mu Q^c P^\mu \pdv{}{Q^a} \bigg\vert_{P} ,
\end{split}
\label{XH}
\eeq
where, in the second line, we have written the vector field in noncanonical coordinates $\{ x^\mu, P_\mu, Q^a \}$, and we have noted explicitly which derivatives are taken keeping $P_\mu$ fixed, as opposed to $p_\mu$.

The collisionless Boltzmann equation for the distribution function $f(x,p,Q)$ can be written in terms of the Liouville vector field as $X_H[f]=0$, and is equivalently known as the Liouville equation. Given the expansion in Eq.~(\ref{fseries}), we require that the Liouville equation be satisfied order by order in $g_s$:
\beq
P^\mu \pdv{f^{(0)}}{x^\mu}\bigg\vert_{P} - \frac{1}{2} \pdv{g^{\alpha \beta}}{x^\mu} \bigg\vert_{P} P_\alpha P_\beta \pdv{f^{(0)}}{P_\mu} = 0
\label{Xf0}
\eeq
and
\beq
\begin{split}
P^\mu \pdv{f^{(1)}}{x^\mu}\bigg\vert_{P} - \frac{1}{2} \pdv{g^{\alpha \beta}}{x^\mu} \bigg\vert_{P} P_\alpha P_\beta \pdv{f^{(1)}}{P_\mu} = Q^a P^\mu F_{\mu\nu}^a \pdv{f^{(0)}}{P_\nu} .
\end{split}
\label{Xf1}
\eeq
We assume that all dependence of $f (x,p,Q)$ on the charges $Q^a$ can only appear at ${\cal O} (g_s)$ or above, and therefore $\partial f^{(0)} / \partial Q^a = 0$. Notice that, as emphasized in Ref.~\cite{Pisarski:1997cp}, the non-Abelian terms in $X_H$, proportional to $g_s f^{abc}$, make no contribution to the evolution of $f^{(1)}$, given our perturbative expansion in $g_s$.

We are free to choose any distribution function for our fluid, as long as it solves the Boltzmann equation and its physical meaning is compatible with small deviations from thermal equilibrium. For this reason, we aim to find a solution of the collisionless Boltzmann equation $X_H[f]=0$ such that the zeroth order in $g_s$ is of the usual form for a canonical ensemble, $f^{(0)}\sim \exp[-\beta E]$, for a constant $\beta$ and a quantity $E$ that may be interpreted as an energy. The equation $X_H [f] = 0$ is typically solved to ${\cal O} (g_s)$ by employing Green's function techniques \cite{Pisarski:1997cp,Blaizot:2001nr,Litim:2001db}. We will instead show that one can further exploit the Hamiltonian formalism to solve efficiently for the distribution function to the same order. The two approaches are equivalent up to ${\cal O} (g_s)$, because quantum corrections to the classical equations of motion in Eq.~(\ref{wongeq}) only arise at ${\cal O} (g_s^2)$ and above.

It is a classic result that Hamiltonian mechanics in the fully covariant phase space $\{ x^\mu, p_\mu, Q^a \}$ with evolution parameterized by an affine $\lambda$ and generated by the Hamiltonian in Eq.~(\ref{Ham}) is equivalent to Hamiltonian mechanics in the reduced phase space $\{ x^i, p_i, Q^a \}$ with evolution parameterized by coordinate time $t \equiv x^0$ and generated by the reduced Hamiltonian
\beq
    \bar{H} \equiv - p_0(t,x^i,p_i,Q^a;h).
\label{Hred}
\eeq
This means that both formalisms yield the same equations of motion. This is possible thanks to the redundancy in the choice of affine parameter $\lambda$, as well as the conservation of $H (x^\mu, p_\mu, Q)$ along phase space trajectories, in the covariant formalism. In Eq.~(\ref{Hred}) we have solved for $p_0$ in terms of $\{t,x^i,p_i,Q^a,h\}$ using $H (x^\mu, p_\mu, Q) \equiv h$, with $h$ a constant. In the covariant formalism, the conservation of $H(x^\mu, p_\mu, Q)$ follows from Eq.~(\ref{wongeq}) and the fact that $H$ cannot depend explicitly on $\lambda$. The reduced phase-space formalism does not yield the conservation of $H(x^\mu,p_\mu,Q^a)$, so we must impose it by hand. For details of the proof, see Sections 44 and 45 in Chapter 9 of Ref.~\cite{Arnold}.

If the reduced Hamiltonian $\bar{H}$ from Eq.~(\ref{Hred}) does not depend explicitly on $t$, then it is also conserved along phase-space trajectories, $X_H [ \bar{H} ] = 0$, so we may exploit $\bar{H}$ to construct a distribution function. That is, we may use the fact that any scalar function of $\bar{H}$ will solve the collisionless Boltzmann equation to devise a valid distribution function for our fluid. We therefore introduce a quasi-stationary approximation: we consider only scenarios in which $\partial_t g_{\mu\nu}$ and $\partial_t A_\mu^a$ remain subdominant. This approximation is appropriate, since we are interested in the behavior of the high-momentum particles, whose dynamics are governed by the time-scale $1/k_{\rm hard} \sim 1 / T$, whereas we expect the soft modes $A_\mu^a (x)$ to evolve on time-scales set by $1 / k_{\rm soft} \gg 1 / k_{\rm hard}$. Likewise, our effective description can only resolve dynamics up to scales set by $1 / k_{\rm soft}$, which bounds how sharply the background spacetime can evolve within our self-consistent expansion as well. In particular, if we allowed $\bar{H}$ to have an explicit dependence on $t$, then 
\beq
X_H [\bar{H}] = \frac{1}{2} \partial_t g^{\mu \nu} P_\mu P_\nu - g_s Q^a P^\mu \partial_t A_\mu^a ,
\label{XHbarH}
\eeq
and thus a distribution function constructed from $\bar{H}$ would yield self-consistent dynamics as long as
\beq
\vert \partial_t g^{\mu\nu} \vert , \frac{ \vert \partial_t A_\mu^a \vert}{\vert A_\mu^a \vert} \ll k_{\rm soft} .
\label{gradients}
\eeq
Eq.~(\ref{gradients}) involves coordinate-dependent quantities. However, any change in $x^\mu$ would be accompanied by changes in the conjugate momenta $p_\mu$, such that Eq.~(\ref{gradients}) remains meaningful, given that $k_{\rm soft}$ is a momentum scale. This follows from the invariance of the phase-space volume under coordinate transformations. As we will see in Section \ref{sec:applications}, this quasi-stationary approximation is easily satisfied in many cosmological applications of interest.

We choose a distribution function $f = \exp [ - \beta_T \bar{H} ]$, where $\beta_T$ is a constant \cite{Sato:2021jaf}. As explained just above, this satisfies the collisionless Boltzmann equation by construction. We will see in the next section that the zeroth-order term $f^{(0)}$ in the $g_s$ expansion corresponds to a canonical ensemble (for fluid undergoing normal flow with respect to coordinate time $t$), so it is a physically reasonable choice because it represents thermal equilibrium. This term is
\beq
f^{(0)} = e^{ \beta_T P_0 } .
\label{f0}
\eeq
Given $f^{(0)}$, we may evaluate the equilibrium occupation numbers per mode in the usual way,
\beq
n_{\rm B, F}^{(0)} \equiv \frac{ \sum n_i e^{ n_i \beta_T  P_0 } }{ \sum e^{ n_i \beta_T  P_0 }} ,
\label{nBFdef}
\eeq
where the sums run from $n_i = 0$ to $n_{i} = \infty$ for bosons and to $n_{i} = 1$ for fermions. Thus we obtain the Bose-Einstein and Fermi-Dirac distributions at zeroth-order:
\beq
    n_{\rm B, F}^{(0)} = \frac{1}{ e^{ - \beta_T  P_0 }\mp 1  } .
\label{nBFfinal}
\eeq
Substituting into Eq.~(\ref{Xf1}) yields
\beq
\begin{split}
n_{\rm B,F}^{(1)} &=  Q^a A_0^a \pdv{ n_{\rm B,F}^{(0)} }{P_0} \\
&= - \beta_T \frac{ e^{-\beta_T P_0}}{(e^{- \beta_T P_0} \mp 1 )^2} Q^a A_0^a 
\label{nBF1}
\end{split}
\eeq
for bosons and fermions, respectively. Summing over species and helicities, we have $\delta f=g_S(2n_{\rm B}^{(1)}+4N_f n_{\rm F}^{(1)})$ in Eq.~(\ref{jdef}). 

To evaluate $j^\mu_a (x)$ we first perform the $Q$ integral with the measure in Eq.~(\ref{dQ}), which yields factors proportional to the index of the representation: $\int dQ \, Q_a Q_b = C_{\rm B,F} \, \delta^a_{\> b}$, with $C_{\rm B}= N_c$ for bosons and $C_{\rm F} = 1/2$ for fermions. Then the current becomes
\beq
\begin{split}
j^\mu_a (x) &=  g_s^2  \int dQ \, d\omega \left(  2 n_{\rm B}^{(1)} + 4 N_f n_{\rm F}^{(1)} \right) Q^a P^\mu \\
&= g_s^2  A_0^a (x) \int d\omega \, {\cal N} (x,P) \, P^\mu ,
\end{split}
\label{jmugeneral}
\eeq
where we have defined
\beq
\begin{split}
{\cal N} (x, P) &\equiv 2 N_c \pdv{ n_{\rm B}^{(0)}}{P_0} + 2 N_f \pdv { n_{\rm F}^{(0)}}{P_0} \\
&= - 2\beta_T \left[  N_c \frac{ e^{-\beta_T P_0}}{(e^{- \beta_T P_0} - 1 )^2} +  N_f \frac{ e^{-\beta_T P_0}}{(e^{- \beta_T P_0} + 1 )^2} \right] .
\end{split}
\label{Ndef}
\eeq
Eq.~(\ref{jmugeneral}) includes contributions from two polarization states for each effectively massless particle, and $2N_f$ fermion species ($N_f$ each for quarks and antiquarks). Note that, as in Minkowski spacetime \cite{Blaizot:2001nr}, $j^\mu_a (x)$ is proportional to $A_0^a (x)$.

\section{Debye Mass}
\label{sec:DebyeMass}

In Minkowski spacetime, the induced current reduces (in the static limit) to $j_\mu^a (x) = m_D^2 A_0^a \delta_\mu^0$, effectively giving a constant mass to the soft gluon components $A_0^a$, which is responsible for Debye screening \cite{Blaizot:2001nr,Kapusta:2006pm,Bazavov:2020teh,Litim:2001db}. In this section, we evaluate the induced current $j^\mu_a (x)$ of Eq.~(\ref{jmugeneral}) and find that in spacetimes of interest, it is proportional to the square of an effective mass, $m_D^2 (x)$, which has spatial dependence.

For any $(3 + 1)$-dimensional, globally hyperbolic spacetime, we can choose coordinates $x^\mu$ to put the metric in the Arnowitt-Deser-Misner (ADM) form (see, e.g., Ref.~\cite{Wald:1984rg})
\beq
\begin{split}
ds^2 &= - N^2 dt^2 + \gamma_{ij} \left( dx^i + \beta^i dt \right) \left( dx^j + \beta^j dt \right) \\
&= - \left( N^2 - \beta_i \beta^i \right)dt^2 + 2 \beta_i \, dt dx^i + \gamma_{ij} dx^i dx^j ,
\end{split}
\label{ds}
\eeq
where $x^0=t$ is a global time function, $N(x)$ is the lapse funtion, and $\beta^i (x)$ the shift vector, whose indices are raised and lowered with $\gamma_{ij}$: $\beta_i = \gamma_{ij} \, \beta^j$. Hypersurfaces of constant time $t$ are Cauchy surfaces by construction, with induced metric $\gamma_{\mu\nu} = g_{\mu\nu} + n_\mu n_\nu$, where 
\beq
n_\mu = - (dt)_\mu = \left( - N, {\bf 0} \right) .
\label{nmu}
\eeq
We 
normalize $t$ such that $N^2\to 1$ on the spatial (possibly asymptotic) boundary. The components of the inverse metric are given by
\beq
g^{00} = - \frac{1}{N^2} , \> g^{0i} = \frac{ \beta^i}{N^2}, \> g^{ij} = \gamma^{ij} - \frac{ \beta^i \beta^j}{N^2} .
\label{inversemetric}
\eeq
In these coordinates, the kinetic momentum takes the form
\beq
P_\mu = \left( -N k + \beta^i P_i , P_i \right) , \>\> P^\mu = \left( \frac{ k}{N} , \gamma^{ij} P_j - \frac{ k}{N} \beta^i \right) , 
\label{Pcomponents}
\eeq
where 
\beq
k \equiv \sqrt{ \gamma^{\mu\nu} \, P_\mu P_\nu} = \sqrt{ \gamma^{ij} \, P_i P_j}
\label{kdef}
\eeq
is the magnitude of the momentum projected on constant-$t$ hypersurfaces.

The spacetime metric $g_{\mu\nu}$ induces a metric in all of phase space, known as the Sasaki metric \cite{Sasaki,Israel1989,Acuna-Cardenas:2021nkj}, such that the invariant momentum volume form is
\beq
d\omega = \frac{ d^4 p}{(2 \pi)^3 \sqrt{ -g}} .
\label{domega1}
\eeq
We restrict to the mass-shell by integrating over $2 \delta (P^2) \Theta (P^0)$. To lowest order in $g_s$, the kinetic and canonical momenta coincide, $P_\mu \rightarrow p_\mu$, so
\beq
\frac{ d^4 p}{ ( 2 \pi)^3 \sqrt{-g}} \rightarrow \frac{ d^3 p_{(i)} }{ (2 \pi)^3 p^0 \, N  \sqrt{\gamma}} ,
\label{domega2}
\eeq
where the last step follows upon noting that $\sqrt{-g} = N \sqrt{ \gamma}$. The current $j^\mu_a (x)$ in Eq.~(\ref{jmugeneral}) may then be written
\beq
\begin{split}
j^\mu_a (x) &= \frac{ g_s^2}{ (2 \pi)^3} \frac{ A_0^a }{N}\\
&\times \int \frac{ d^3 p_{(i)}}{\sqrt{\gamma }}  {\cal N} (x, p) \left[ \delta^\mu_{0} + \delta^\mu_i \left( \beta^i - \frac{ N}{k} \gamma^{ij} p_j \right) \right].
\label{j1}
\end{split}
\eeq
We next consider a change in momentum coordinates $\tilde{k}^i \equiv ( \gamma^{1/2} )^{ij} p_j$, where $(\gamma^{1/2} )^{ij}$ is the square matrix that satisfies $(\gamma^{1/2} )_{mi} \gamma^{ij} (\gamma^{1/2} )_{j \ell} = \delta_{m \ell}$. Then $d^3 p_{(i)} = \sqrt{\gamma} \, d^3 \tilde{k}^{(i)}$, and $k^2 = \tilde{k}^2 = \delta_{ij} \tilde{k}^i \tilde{k}^j$, which is even in the components $\tilde{k}^i$. One could perform an additional coordinate transformation as in Ref.~\cite{Brandt:1994mv}, $p_i' = p_i - \beta_i p_0 / g_{00}$, to aid in evaluating the integral in Eq.~(\ref{j1}). For the applications of interest to us, we will instead restrict attention to spacetimes in which the shift vector vanishes, $\beta^i = 0$. Then the integration in Eq.~(\ref{j1}) may be performed exactly. In such cases, we find
\begin{align}
j^0_a (x) &= \frac{ g_s^2}{ (2 \pi)^3} \frac{ A_0^a}{N} \int d^3 \tilde{k}^{(i)} \, {\cal N} (x, \tilde{k} ) , \label{j0a} \\
j^i_a (x) &= - \frac{ g_s^2}{(2 \pi)^3} A_0^a (\gamma^{1/2})^{ij} \delta_{j\ell} \int d^3 \tilde{k}^{(i)} \, {\cal N} (x, \tilde{k}) \frac{  \tilde{k}^\ell }{k} \label{jia}.
\end{align}
The quantity ${\cal N} (x, \tilde{k})$ is even in the components $\tilde{k}^i$, so that $j^i_a (x)$ vanishes identically. We further note that the momentum coordinates $\tilde{k}^i$ are Cartesian, so $j^0_a (x)$ is equivalent to
\beq
\begin{split}
j^0_a (x) &=  \frac{ g_s^2}{ 2 \pi^2} \frac{ A_0^a}{N} \int_0^\infty d \tilde{k} \, \tilde{k}^2 \, {\cal N} (x, \tilde{k} ) \\
&= - \frac{ m_D^2 (x) }{N^2 (x)} A_0^a (x) ,
\end{split}
\label{j0final}
\eeq
with
\beq
m_D^2 (x) \equiv \frac{1}{6} \left( 2 N_c + N_f \right) \frac{ g_s^2}{(\beta_T N(x) )^2} .
\label{mD}
\eeq 
Eq.~(\ref{j0final}) corresponds to
\beq
j_\mu^a (x) = m_D^2 (x) A_0^a \, \delta_\mu^0 .
\label{jlower}
\eeq
The induced current in Eq.~(\ref{jlower}) is equivalent to inserting an effective mass $m_D (x)$ for the $A_0^a$ soft gluon components within the effective action of Eq.~(\ref{Seff}).

The expression for $m_D (x)$ in Eq.~(\ref{mD}) reduces to the usual expression in Minkowski spacetime upon setting $ N (x) \rightarrow 1$ and identifying $1/\beta_T = T$ with the temperature of the plasma \cite{Blaizot:2001nr,Kapusta:2006pm,Bazavov:2020teh,Weldon:1982aq,Rebhan:1993az,Rebhan:1994mx,Braaten:1995jr,Boyanovsky:1999jh,Blaizot:1995kg,Litim:2001db,Mrowczynski:2016etf}. In our case, the Boltzmann factor for the high-momentum particles is $\exp [ - \beta_T \bar{H}^{(0)}] = \exp [ \beta_T N(x) k (x) ]$, since we are considering spacetimes in which $\beta^i = 0$. This is exactly what one would expect for particles in a fluid that is undergoing normal flow, with four-velocity $n^\mu (x)$ and therefore local energy per particle $E \equiv - n^\mu p_\mu = k$, if we identify $T (x) \equiv 1 / ( \beta_T N(x))$ with the local temperature. We may therefore identify the constant $\beta_T \equiv 1 / T_0$ and write
\beq
T (x) = \frac{ T_0}{N(x)} .
\label{Tx}
\eeq
Given that we have normalized $t$ such that $N (x) \rightarrow 1$ on the spatial (possibly asymptotic) boundary, $T_0$ is the temperature associated with time $t$ on the spatial boundary. We have thus recovered the familiar Tolman temperature gradient in a curved spacetime \cite{Tolman:1930ona,Tolman:1930zza,Santiago:2018lcy}. The Debye mass then takes the form
\beq
m_D^2 (x) = \frac{1}{6} \left( 2 N_c + N_f \right) g_s^2 \, T^2 (x),
\label{mDfinal}
\eeq
with the local temperature $T (x)$ given by Eq.~(\ref{Tx}).

\section{Cosmological Applications}
\label{sec:applications}

In this section we consider a few specific examples. We begin with the familiar case of Debye screening of a non-Abelian plasma in Minkowski spacetime, to identify several regimes of interest. Next we generalize to the case of a spatially flat Friedmann-Lem\^{a}itre-Robertson-Walker (FLRW) spacetime, which introduces a new scale (compared to the Minkowski case), set by the Hubble radius. In the last subsection, we examine Debye screening in the vicinity of a primordial black hole.

\subsection{Debye screening in Minkowski spacetime}
\label{sec:DebyeMinkowski}

We first consider the behavior of soft gluon modes $A_\mu^a (x)$ within a hot plasma in Minkowski spacetime, to clarify notation and identify several physical regimes of interest. In that case, the lapse function becomes trivial, $N (x) = 1$, and the shift vector vanishes, $\beta^i = 0$. In the static limit, $\partial_0 A_\mu^a (x) = 0$, the exact equations of motion of Eq.~(\ref{Aeom2}) reduce to
\beq
\begin{split}
    \delta_\nu^{0} &\bigg\{ \partial_i F^a_{i0} + g_s f^{abc}  A_i^b F_{i0}^c \bigg\} \\
    & + \delta_{\nu}^{i} \bigg\{ \partial_k F_{k i}^a + g_s f^{abc} \left( A_0^b F_{i0}^c + A_k^b F_{k i}^c \right) \bigg\} = j_\nu^a ,
\end{split}
\label{eomM1}
\eeq
where repeated indices are summed and we have adopted Cartesian coordinates for the spatial sections. The induced current of Eq.~(\ref{jlower}) reduces to
\beq
j_\nu^a ({\bf x}) = m_D^2 A_0^a ({\bf x}) \delta_\nu^{\>\> 0} + {\cal O} (g_s^3 ) .
\label{jmuMink}
\eeq

A generic feature of non-Abelian field theories, which has been well-studied for the case of self-interacting gauge fields in Minkowski spacetime, is the existence of monopole-like solutions among the spatial components $A_i^a ({\bf x})$. Such solutions can be found in simplest form for SU(2) \cite{Wu:1976qk,Wu:1976ge,Jackiw:1993pc,Blaizot:2001nr}, and have been generalized to monopoles charged under various 3-dimensional subgroups of SU$(N_c)$ \cite{Sinha:1976bw,Preskill:1984gd,Kunz:1987ef}. Moreover, it is well-known that self-consistent, static solutions to the non-Abelian equations of motion in Minkowski spacetime generically include a Yukawa-like screened behavior for the component $A_0^a ({\bf x})$ combined with the Wu-Yang monopole solution for the components $A_i^a ({\bf x})$ \cite{Blaizot:2001nr,Jackiw:1993pc}. Such solutions underscore the important physical point that only the components $A_0^a ({\bf x})$ acquire a nonzero mass within a medium in the state described in Section \ref{sec:DebyeMass}---as indicated by the form of the induced current $j_\mu^a ({\bf x})$ in Eq.~(\ref{jmuMink})---and hence only those components undergo screening, with amplitude proportional to $\exp [ - m_D r]$. 

Our aim in this section is to examine how well-known field configurations in Minkowski spacetime generalize to various curved spacetimes of interest, in which additional length-scales become relevant. We therefore begin by considering an exact solution to Eq.~(\ref{eomM1}) that consists of a superposition of a screened component $A_0^a$ with a Wu-Yang monopole solution for the components $A_i^a$. For the case of ${\rm SU} (2)$, the solution takes the form \cite{Wu:1976qk,Wu:1976ge,Jackiw:1993pc,Blaizot:2001nr}
\beq
A_0^a ({\bf x}) = -\frac{ {\cal Q}^a_0  \, e^{- m_D r}}{r} , \>\> A_i^a ({\bf x}) = \frac{ \epsilon_{aij} \hat{x}^j }{g_s r} ,
\label{WuYang}
\eeq
where ${\cal Q}^a_0 = {\cal Q}_0 \,\hat{x}^a$, $Q_0$ is a constant, $\hat{x}^j \equiv x^j / r$ is a unit vector, and $\epsilon_{ijk}$ is the usual Levi-Civita symbol. Note that the solution mixes spatial indices and color-space indices, which is straightforward for ${\rm SU} (2)$, since both $i, j = 1, 2, 3$ and $a, b = 1, 2, 3$. (To confirm that $A_\mu^a ({\bf x})$ in Eq.~(\ref{WuYang}) satisfies Eq.~(\ref{eomM1}), it is helpful to construct the projection operator $P^a_{\>\>b} = \delta^a_{\>\> b} - \hat{x}^a \hat{x}_b$, noting that $\hat{x}_a P^a_{\>\> b} = 0 = \hat{x}^b P^a_{\>\> b}$ and $\partial_j \hat{x}^i = r^{-1} P^i_{\>\> j}$.) For SU(3), exact monopole solutions have been found for subgroups such as ${\rm U}(1) \times {\rm U} (1)$ and U(2), for which the components $A_i^a$ have comparable asymptotic behavior to the solution in Eq.~(\ref{WuYang}) \cite{Sinha:1976bw,Preskill:1984gd,Kunz:1987ef}.

The solution for $A_\mu^a ({\bf x)}$ in Eq.~(\ref{WuYang}) carries both chromoelectric charge ${\cal Q}$ and chromomagnetic charge ${\cal P}$. We consider the quasi-local charges defined in terms of the chromoelectric and chromomagnetic fields, $E^a_i \equiv F_{0i}^a$ and $B^a_i \equiv -\frac{1}{2} \epsilon_{ijk} F^{jk}_a$, respectively. Specifically, we identify charges via the relations
\beq
    E^a_i E^a_i = \frac{{\cal Q}^2(r)}{r^4},\>\>
    B^a_i B^a_i = \frac{{\cal P}^2(r)}{r^4},
\label{EB}
\eeq
where the spatial and color indices are summed over. Note that the fields $E_i^a$ and $B_i^a$ transform covariantly in color space under gauge transformations, and therefore the bilinear combinations in Eq.~(\ref{EB}) are gauge-invariant. For the solution given in Eq.~(\ref{WuYang}), the charges are
\beq
{\cal Q} (r) = {\cal Q}_0 \left(1 + m_D r \right) e^{- m_D r} , \>\> {\cal P} (r) = \frac{1}{g_s} ,
\label{QPr}
\eeq
where ${\cal Q}_0 = \sqrt{{\cal Q}_0^a{\cal Q}_0^a}$ is the chromoelectric charge measured at the origin. Because of screening, the value of ${\cal Q} (r)$ measured at a finite distance from the origin will be reduced compared to ${\cal Q}_0$. The chromomagnetic charge, on the other hand, is not screened, and hence ${\cal P} = 1 / g_s$ everywhere. 

Whereas the magnitude of the chromomagnetic charge ${\cal P}$ is fixed by $1 / g_s$, the chromoelectric charge ${\cal Q}_0$ at the origin is a free parameter and can be arbitrarily larger than $1/ g_s$. Such a scenario would be compatible with a collection of test charges at the origin such that the energy associated with the charges does not backreact on the spacetime itself. In that case, there will exist a self-consistent regime, $0 \leq r \leq r_{\rm abel}$, within which $\vert A_0^a ({\bf x}) \vert \gg \vert A_i^a ({\bf x}) \vert$, with $r_{\rm abel}$ given by
\beq
r_{\rm abel} \equiv \frac{1}{ m_D} {\rm ln} \left[  g_s {\cal Q}_0 \right] .
\label{rmagdef}
\eeq
For $0 \leq r \leq r_{\rm abel}$, the system is quasi-Abelian, with $A_\mu^a ({\bf x}) \simeq A_0^a ({\bf x}) \delta_\mu^{\>\> 0}$, and hence non-Abelian contributions of the form $f^{abc} A_\mu^a A_\nu^b$ remain subdominant. Within this regime, one may solve the exact equations of motion of Eqs.~(\ref{eomM1})--(\ref{jmuMink}) with the ansatz $A_\mu^a({\bf x})=A_0^a({\bf x})\delta^0_\mu$. The solution for $A_\mu^a ({\bf x})$ is then of the form in Eq.~(\ref{WuYang}), but with $A_i^a ({\bf x}) = 0$
and ${\cal Q}^a_0$ an arbitrary, constant vector in color space.

\subsection{Debye screening in an FLRW spacetime}
\label{sec:DebyeFLRW}

We next consider Debye screening within a hot quark-gluon plasma in an expanding FLRW spacetime at early times, prior to the QCD confinement transition ($t < t_{\rm QCD} \sim 10^{-5}\,{\rm s}$). The line-element for a spatially flat FLRW universe may be written
\beq
ds^2 = - dt^2 + a^2 (t) \, \left[ dr^2 + r^2 d\Omega_{(2)}^2 \right],
\label{dsFLRW}
\eeq
where $a(t)$ is the scale factor and $r$ is a comoving radial coordinate. The Hubble parameter is given by $H(t) \equiv \dot{a} / a$, where overdots denote derivatives with respect to cosmic time $t$, and the Hubble radius is $r_H = 1 / H$. The presence of $r_H$ introduces a new scale compared to Minkowski spacetime, so now we must consider $H$ as well as $k_{\rm hard}$ and $k_{\rm soft}$. Note also that in these coordinates the lapse function is $N (x) = 1$, so the temperature has no spatial gradients.

The term ${\cal L}_{\rm fluid}$ in the effective action of Eq.~(\ref{Seff}) includes contributions from the high-momentum quarks and gluons in the QGP, which, as noted in Section \ref{sec:current}, behave to leading order as a gas of non-interacting particles with a radiation-dominated equation of state. We assume that the (coarse-grained) energy density $\rho$ and pressure $P$ associated with the high-momentum particles dominate $T_{\mu\nu}$, so that the Friedmann equation takes the form
\beq
H^2 = \frac{1}{ 3 M_{\rm pl}^2} \left( \frac{ \pi^2}{30} g_* T^4 \right)
\label{Friedmann}
\eeq
corresponding to a fluid of $g_*$ effectively massless degrees of freedom in equilibrium at temperature $T$; for the Standard Model at temperatures much greater than the top-quark mass ($m_t = 173 \, {\rm GeV}$), $g_* = 106.75$ \cite{Mukhanov:2005sc,Boyanovsky:2006bf}. During the radiation-dominated phase, $a(t) \propto t^{1/2}$, so we have $H(t) = 1 / (2t)$ and 
\beq
T (t) = \left( \frac{ 90}{\pi^2 g_*} \right)^{1/4} \left( \frac{ M_{\rm pl}}{2t} \right)^{1/2} ,
\label{Tflrw}
\eeq
consistent with adiabatic expansion, $T (t) \, a (t) = {\rm constant}$.

The equations of motion from Eq.~(\ref{Aeom2}) are
\beq
\begin{split}
j_\nu^a &= \frac{1}{a^2}\,\delta_\nu^{0} \bigg\{ \partial_i F^a_{i0} + g_s f^{abc} A_i^b F_{i0}^c \bigg\}\\
&+\delta_{\nu}^{i} \bigg\{ HF_{i0}^a + \partial_0 F_{i0}^a + \frac{1}{a^2} \partial_k F_{k i}^a\\
&\qquad \qquad + g_s f^{abc} \left( A_0^b F_{i0}^c + \frac{1}{a^2} A_k^b F_{k i}^c \right) \bigg\},
\end{split}
\label{eomFLRW}
\eeq
where repeated indices are summed. The Hubble parameter $H(t)=\dot{a}/a$ sets the time-scale over which we expect cosmological dynamics to be relevant. Notice that the terms that make Eq.~(55) differ from Eq.~(46) (up to factors of $a(t)$) are either proportional to $H(t)$ or include a time derivative $\dot{A}_\mu^c$. We consider an ansatz for $A_\mu^c (x)$ in which the only time dependence arises from the scale factor $a(t)$. By the chain rule, $\dv*{A_\mu^c}{t} = \dot{a}\pdv*{A_\mu^c}{a}=H\pdv*{A_\mu^c}{ \,{\rm ln} \,a}$. Therefore any term with a time derivative $\dot{A}_\mu^c$ will be proportional to $H$. If $H\ll k_{\rm soft}^2/k_{\rm hard} \sim g_s^2T$, then all terms proportional to $H$, which include those with time derivatives $\dot{A}_\mu^c$, will be subleading compared to $j_\nu^c\sim m_D^2A_0^c\sim k_{\rm soft}^2k_{\rm hard}$.

If the hierarchy $H\ll g_s^2T$ is satisfied at early times, it will be even more easily satisfied at later times, since $H\sim T^2/M_{\rm pl}$ from Eq.~(\ref{Tflrw}) and $T/M_{\rm pl}$ decreases as the universe cools down, while $g_s$ gradually runs to larger values. Now we proceed to show that $H\ll g_s^2T$ holds at early times of interest.

Consider, as an example, dynamics at the time $t_c \sim 10^{-21} \, {\rm s}$, the earliest time that a population of primordial black holes (PBHs) could have formed, if the PBHs are to constitute all of dark matter today \cite{Carr:2020xqk,Green:2020jor,Villanueva-Domingo:2021spv,Escriva:2022duf}. At such early times, the Hubble scale $H (t_c) = 3.0 \times 10^{-4} \, {\rm GeV}$ and the fluid filling the FLRW spacetime had a temperature $T (t_c) = 10^7 \, {\rm GeV}$. At such high energy scales, we must take into account the running of the strong coupling $\alpha_s \equiv g_s^2 / (4 \pi)$. To two-loop order, the running with energy scale $\mu$ is given by \cite{Bazavov:2020teh}
\beq
\dv{ \alpha_s (\mu) }{ \, {\rm ln} \mu} = - 2 \alpha_s \left[ b_0 \frac{ \alpha_s}{4 \pi} + b_1 \left( \frac{ \alpha_s}{4 \pi} \right)^2 \right] ,
\label{alpharun}
\eeq
with
\beq
\begin{split}
b_0 &= \frac{ 11}{3} N_c - \frac{2}{3} N_f , \\
b_1 &= \frac{32}{3} N_c^2 - \frac{10}{3} N_c N_f - \left( \frac{ N_c^2 - 1}{N_c} \right) N_f .
\end{split}
\label{b0b1}
\eeq
Upon normalizing $\alpha_s (m_Z) = 0.118$ at the $Z$ boson mass ($m_Z = 91.2 \, {\rm GeV}$), we find $\alpha_s = 0.046$ at the energy scale $\mu = T (t_c) = 10^7$ GeV. This yields a ratio $g_s^2T/H\sim 10^9$.
Hence we may self-consistently neglect $H \ll g_s^2T$ when considering the dynamics of the gluon modes on length-scales $\lambda_D$, even at very early times in cosmic history. In other words, the FLRW spacetime is consistent with our quasi-stationary approximation from Eq.~(\ref{gradients}), since $|\partial_tg^{\mu\nu}|\sim H \ll g_sk_{\rm soft}<k_{\rm soft}$.

Having shown that we may neglect cosmological dynamics, the scale factor $a(t)$ will be approximately constant in regimes of interest. We may therefore rescale the spatial coordinates as $x^i\rightarrow x^i/a$, and thus recover identical equations of motion to those in the Minkowski background, Eq.~(\ref{eomM1}). A solution is the superposition of a screened component $A_0^a$ with a Wu-Yang monopole for the components $A_i^a$, which in the case of ${\rm SU}(2)$ takes a similar form to Eq.~(\ref{WuYang}),
\beq
A_0^a (x) = -\frac{ {\cal Q}_0^a  \, e^{- m_D a r}}{ar} , \>\> A_i^a (x) = \frac{ \epsilon_{aij} \hat{x}^j }{g_s ar} ,
\label{WuYangFLRW}
\eeq
where we have rescaled the spatial coordinates back so that $r$ is the comoving radial coordinate from Eq.~(\ref{dsFLRW}). The solution for $A_\mu^a (x)$ in Eq.~(\ref{WuYangFLRW}) remains consistent with the quasi-stationary approximation of Eq.~(\ref{gradients}) for $\lambda_D \leq r \ll r_H$.

\subsection{Debye screening near a primordial black hole}
\label{sec:DebyePBH}

Our final example concerns Debye screening within the hot plasma surrounding a primordial black hole (PBH) that formed early in cosmic history. PBHs form via direct gravitational collapse of primordial overdensities, and their masses are typically proportional to the Hubble mass $M_H$ at the time of formation $t_c$, where $M_H (t_c) = 4 \pi M_{\rm pl}^2 / H(t_c)$ is the mass contained within a Hubble radius $H^{-1} (t_c)$. Such black holes may therefore form with a huge range of masses, depending on their time of formation \cite{Carr:2020xqk,Green:2020jor,Villanueva-Domingo:2021spv,Escriva:2022duf}.

Any PBHs that formed at times $t < t_{\rm QCD} = 10^{-5} \, {\rm s}$ would undergo collapse amid a hot plasma of unconfined quarks and gluons. On long length-scales at such early times, $\lambda \gg 1 / k_{\rm soft}$, we expect that the plasma would have attained a charge-neutral equilibrium distribution. Yet on shorter length-scales, set by $\lambda_D = 1 / m_D \sim 1 / k_{\rm soft} \sim 1 / (g_s T)$, spatial regions of nonvanishing net color charge can form, 
within which the charges for most soft gluons align in color space
\cite{Manuel:2003zr,Manuel:2004gk}. In such scenarios, the PBHs would form by absorbing one or more net-color regions. Depending on the ratio of the PBH radius ($\sim G M_{\rm PBH}$) and the Debye length ($\sim 1/m_D$) at the time of formation, PBHs therefore could form with net color charge ${\cal Q}_0$. Such a scenario is distinct from the examples reviewed in Ref.~\cite{Volkov:1998cc,Volkov:2016ehx}, which concern black hole solutions to the Einstein-Yang-Mills equations in vacuum. In our case, the PBHs are immersed in a hot, active medium.

We consider a scenario in which a PBH forms with a small net charge ${\cal Q}_0$; the case of larger ${\cal Q}_0$ is treated in Ref.~\cite{Alonso-Monsalve:2023brx}. In particular, we consider the regime
\beq
{\cal Q}_0^2 \ll G M_{\rm PBH}^2 .
\label{Qsmall}
\eeq
Within this regime, backreaction on spacetime from the energy associated with the enclosed  charge ${\cal Q}_0$ remains subdominant, and we may consider the dynamics of soft gluon modes $A_\mu^a (x)$ within a fixed background geometry. (For ${\cal Q}_0^2 \sim G M_{\rm PBH}^2$, one must solve the Einstein field equations as well as the equations of motion for $A_\mu^a (x)$, which we consider in separate work \cite{EAMprep}.) 

We consider a spherically symmetric spacetime with a black hole of mass $M_{\rm PBH}$ at the origin surrounded by hot plasma with a radiation-dominated equation of state. Our formalism holds in a quasi-static limit, which requires that $M_{\rm PBH}$ not change appreciably over time-scales set by $1/k_{\rm soft}$. Two competing effects could change $M_{\rm PBH}$: evaporation due to Hawking radiation (which would reduce $M_{\rm PBH}$ over time) and accretion from the surrounding medium (which would increase $M_{\rm PBH}$ over time). For the regime of interest, we find that both of these effects are negligible over the relevant dynamical time-scale, and hence we may neglect $\dot{M}_{\rm PBH}$.

Consider first evaporation from Hawking radiation. Because we are considering PBHs with modest charge, subject to the inequality of Eq.~(\ref{Qsmall}), we may approximate the Hawking temperature based on that for an (uncharged) Schwarzschild black hole of mass $M_{\rm PBH}$. (Any net enclosed charge would decrease the black hole's surface gravity, and hence its Hawking temperature, compared to the zero-charge case, thus rendering evaporation even less efficient.) The Hawking temperature for a Schwarzschild black hole is given by \cite{Wald:1984rg}
\beq
T_H = \frac{ M_{\rm pl}^2}{M_{\rm PBH}}.
\label{THdef}
\eeq
The typical mass for a PBH is set by the mass enclosed within a Hubble volume $M_H (t_c)$ at the time of collapse $t_c$ \cite{Carr:1975qj,Niemeyer:1997mt,Green:1999xm,Kuhnel:2015vtw}:
\beq
M_{\rm PBH} (t_c) = \gamma M_H (t_c) = 4 \gamma \sqrt{ \frac{90}{g_*}} \left( \frac{ M_{\rm pl}}{T_c} \right)^2 M_{\rm pl},
\label{barM}
\eeq
where $\gamma \simeq 0.2$ and $T_c$ is the temperature of the plasma at the time $t_c$. The values of $M_{\rm PBH}$ relevant to account for dark matter lie within the range $10^{17} \, {\rm g} \leq M_{\rm PBH} \leq 10^{22} \, {\rm g}$ \cite{Carr:2020xqk,Green:2020jor,Villanueva-Domingo:2021spv,Escriva:2022duf}; from Eq.~(\ref{barM}), these correspond to plasma temperatures $10^5 \, {\rm GeV} \leq T_c \leq 10^7 \, {\rm GeV}$. Meanwhile, for masses within the dark-matter range, Eq.~(\ref{THdef}) yields $10^{-9} \, {\rm GeV} \leq T_H \leq 10^{-4} \, {\rm GeV}$, exponentially lower than the temperature of the surrounding plasma. Such PBHs will therefore be net absorbers (rather than emitters) around the time of their formation.

Next we may consider accretion. One might suppose that if the fluid were moving with respect to the black hole, $M_{\rm PBH}$ would increase by absorbing some of the nearby fluid. Nonetheless, even if the relative speed between the fluid and the black hole approached the speed of light, and the black hole absorbed all the fluid in its path, then $M_{\rm PBH}$ would only increase by a fraction $\Delta M / M_{\rm PBH} \sim {\cal O} (1) \, {\cal R} \, T / M_{\rm pl}$ over an entire Hubble time, where ${\cal R}$ is the ratio of the black hole radius to the Debye screening length and $T$ the temperature of the plasma. As in the previous section, we are interested in early times at which the plasma temperature could have been as high as $T \sim 10^7$ GeV, in which case $T / M_{\rm pl} \sim 10^{-11}$. We consider black holes for which the ratio ${\cal R}$ is not exponentially larger than one, so that the PBH forms with some small, residual color charge, and therefore $\Delta M / M$ remains negligible. A more formal calculation of the Bondi accretion rate for PBHs in this scenario yields the same conclusion \cite{ShapiroTeukolsky1983,Rice:2017avg,DeLuca:2020fpg,Richards:2021zbr}.

Within the regime indicated in Eq.~(\ref{Qsmall}), the spacetime can be described by the McVittie line-element, which reduces to an ordinary Schwarzschild spacetime near the origin and asymptotes to a spatially flat FLRW spacetime at large distances \cite{Kaloper:2010ec,Faraoni:2014nba,Gaur:2022hap}. A convenient parameterization may be written \cite{Gaur:2022hap} 
\beq
\begin{split}
ds^2 &= - f (t, r) \, dt^2 \\
&\quad + \left[ \frac{ a(t) \left\{ dr + H (t) \, r dt \right\} }{\sqrt{ f (t,r)}}  - H(t) \, a (t) \, r dt\right]^2 \\
&\quad + a^2 (t) \, r^2 d\Omega_{(2)}^2 ,
\end{split}
\label{dsMcVittie}
\eeq
with
\beq
f (t, r) \equiv 1 - \frac{ r_s}{a (t) r}
\label{ftr}
\eeq
and $r_s \equiv 2 G M_{\rm PBH}$ the usual Schwarzschild radius. 

We found in Section \ref{sec:DebyeFLRW} that in cosmological scenarios of interest, $H \ll m_D$, even at very early times. For the remainder of this section, we will therefore set $H \sim 0$, for which $a (t) \sim {\rm constant}$ (which we will scale to 1). In that limit, the lapse function reduces to $N(x) \rightarrow \sqrt{ f (r)}$, with $f (r) = 1 - r_s / r$, and the shift vector vanishes, $\beta^i \rightarrow 0$.

Next we consider an appropriate range for the enclosed charge ${\cal Q}_0$, subject to the constraint in Eq.~(\ref{Qsmall}). The shortest scales that can be resolved within our EFT are set by $\lambda_D = 1 / m_D$, so we restrict $r_s \geq \lambda_D$, which in turn requires $M_{\rm PBH} \geq 1 / (2 G m_D)$. Then Eq.~(\ref{Qsmall}) corresponds to the upper bound
\beq
{\cal Q}_0 \ll \frac{ 1}{\sqrt{ 2 \alpha_s}} \, \frac{ M_{\rm pl}}{T}  \sim 10^{12} 
\label{Qsmall2}
\eeq
around $t_c \sim 10^{-21} \, {\rm s}$. Meanwhile, as discussed around Eq.~(\ref{rmagdef}), if ${\cal Q}_0 \gg 1 / g_s$, then the soft modes $A_\mu^a$ will assume a quasi-Abelian form, with $\vert A_0^a \vert \gg \vert A_i^a \vert$, for $r < r_{\rm abel}$. The estimate of $r_{\rm abel}$ in Eq.~(\ref{rmagdef}) holds in homogeneous spacetimes, for which the temperature has no spatial gradients. In the present case, given Eq.~(\ref{Tx}), $m_D (x) \sim g_s T_0 / \sqrt{ f (r)}$. Following the same steps that led to Eq.~(\ref{rmagdef}), we find
\beq
\frac{ r_{\rm abel}}{r_s} = \frac{1}{2} \left[ 1 + \sqrt{ 1 + \left( \frac{ 2 \tilde{\lambda}_D}{r_s} {\rm ln} [ g_s {\cal Q}_0 ] \right)^2 } \right] ,
\label{rabelSch}
\eeq
where $\tilde{\lambda}_D$ is the Debye length associated with the temperature $T_0$. For $r_s / \tilde{\lambda}_D = 3.5$ and ${\cal Q}_0 \sim 10^4$, this yields $r_{\rm abel} \sim 3 \, r_s$, while ${\cal Q}_0 \sim 10^{10}$ yields $r_{\rm abel} \sim 7 \, r_s$.

\begin{figure}
    \centering
    \includegraphics[width=3in]{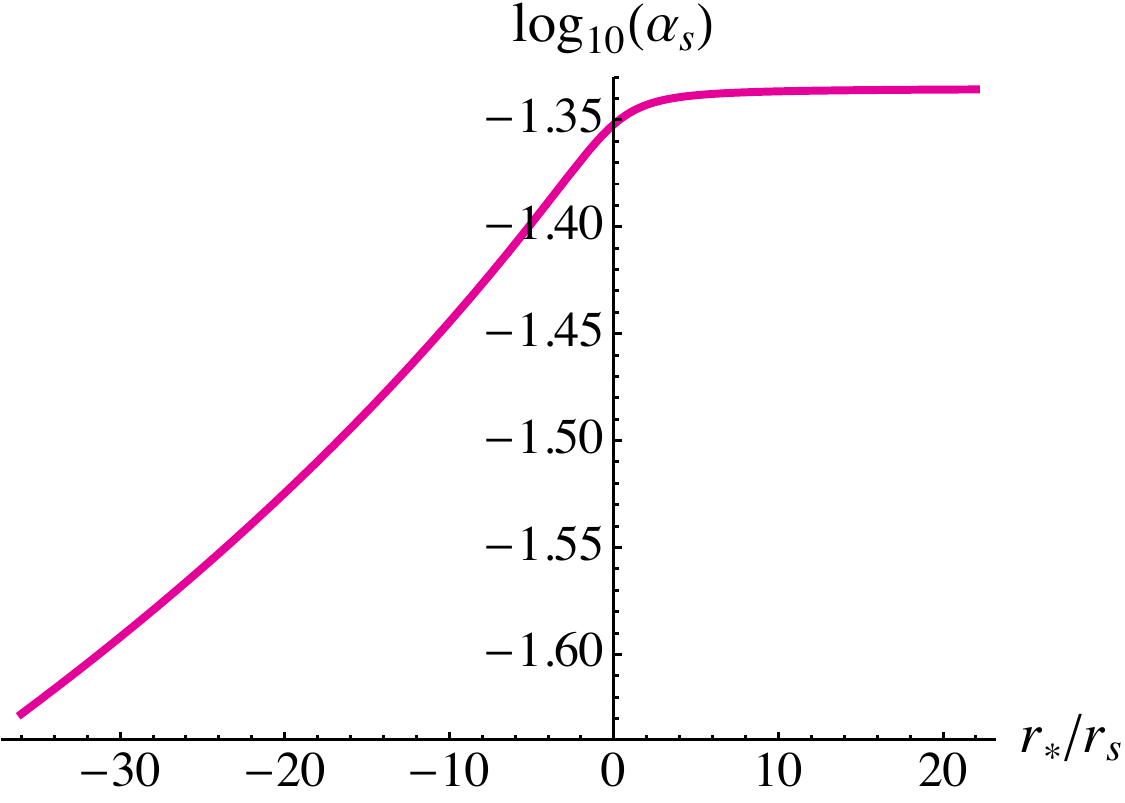}
    \caption{The QCD coupling strength $\alpha_s \equiv g_s^2 / (4 \pi)$ runs to lower values near the event horizon of a black hole, as the temperature of the surrounding plasma increases. Here $r_*$ is the ``tortoise" radial coordinate, $r_* \equiv r + r_s {\rm ln} [ ( r / r_s) - 1]$; $r_* / r_s = -36$ corresponds to $r / r_s = 1 + 10^{-16}$. We have set $T_0 = 10^7\, {\rm GeV}$ on the asymptotic boundary.}
    \label{fig:alpha}
\end{figure}

The last departure to consider from the previous examples concerns the effect of the local temperature $T (x)$ on the coupling $\alpha_s$. Since the local temperature $T(r)$ increases as $r$ approaches $r_s$, the local QCD strength runs toward zero as one approaches the black hole event horizon: an example of asymptotic freedom within an inhomogeneous spacetime. To quantify the effect, we use Eq.~(\ref{alpharun}) for the running of $\alpha_s$ with energy scale $\mu \rightarrow T (r)$. To resolve the behavior of $\alpha_s$ near $r_s$, we adopt the ``tortoise" radial coordinate, $r_* \equiv r + r_s {\rm ln} [ ( r / r_s) - 1]$, for which $r = r_s$ corresponds to $r_* \rightarrow - \infty$ \cite{Wald:1984rg}. See Fig.~\ref{fig:alpha}.

\begin{figure}
    \centering
    \includegraphics[width=3in]{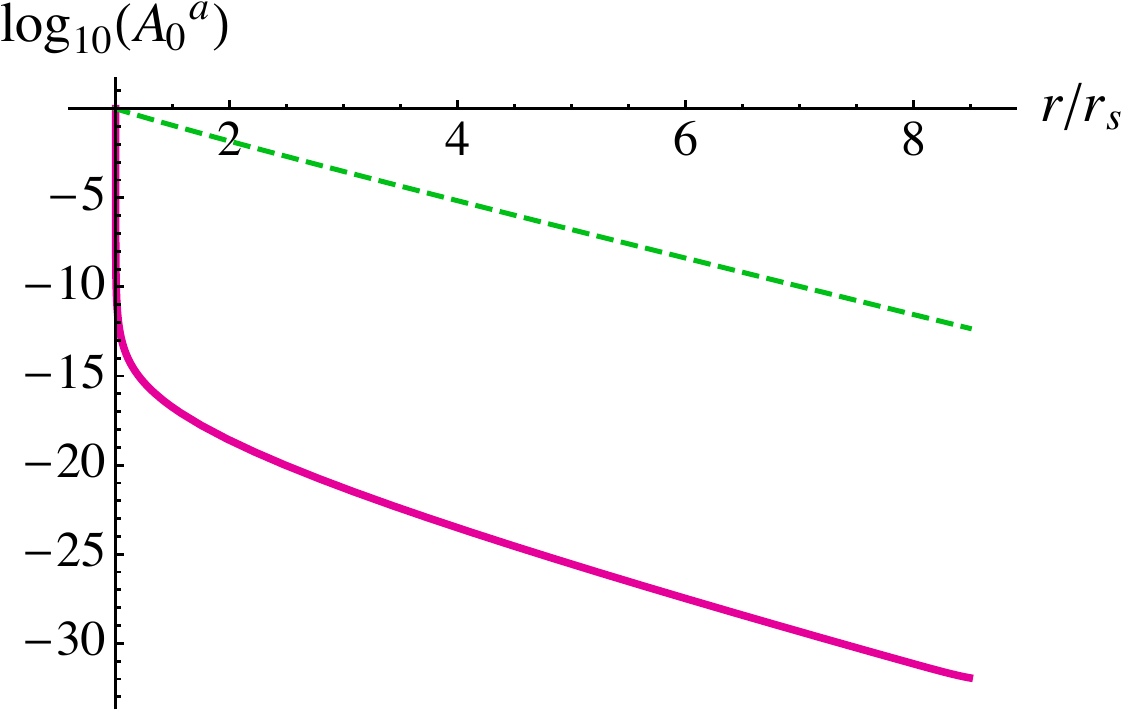}
    \caption{The soft mode $A_0^a (r)$ (solid line) for $r_s / \tilde{\lambda}_D = 3.5$ and $T_0 = 10^7 \, {\rm GeV}$. Given the greater density of plasma near the black hole, Debye screening is particularly effective for $r \gtrsim r_s$. For $r \gg r_s$, the behavior of $A_0^a (r)$ asymptotes to the expected slope, $\exp [ - m_D r ] / r$ (dashed line). }
    \label{fig:A0}
\end{figure}

In the vicinity of a primordial black hole, the evolution of the soft modes $A_\mu^a (x)$ is therefore characterized by a hierarchy of length-scales:
\beq
\lambda_D < r_s < r_{\rm abel} \ll r_H .
\label{lengthscales}
\eeq
For $r_s \leq r \leq r_{\rm abel}$, the soft modes will evolve as a quasi-static, quasi-Abelian system within a fixed background spacetime. In that regime, $A_\mu^a (x) \simeq A_0^a (r) \delta_\mu^0 + {\cal O} (r / r_{\rm abel})$, and the equations of motion in Eq.~(\ref{Aeom2}) reduce to
\beq
\partial_r^2 A_0^a (r) + \frac{2}{r} \partial_r A_0^a (r) - \frac{ \tilde{m}_D^2 }{f^2 (r)} A_0^a (r) = 0 ,
\label{eomPBH}
\eeq
where $\tilde{m}_D = 1 / \tilde{\lambda}_D$ is the Debye mass associated with the temperature $T_0$. Including the running of $g_s$ with $r$, we may solve Eq.~(\ref{eomPBH}) numerically, with a typical example shown in Fig.~\ref{fig:A0}. The component $A_0^a (r)$ undergoes strong screening for $r \gtrsim r_s$, since more plasma gathers near the event horizon, yielding a higher density and temperature than at locations $r \gg r_s$. Far from the black hole, $A_0^a (r)$ asymptotes to similar behavior as found in Sections \ref{sec:DebyeMinkowski} and \ref{sec:DebyeFLRW}, with $A_0^a \sim \exp[ - m_D r ] / r$. Note that an observer at $r \gtrsim r_s$ would measure an effective charge ${\cal Q} (r)$ much smaller than the charge ${\cal Q}_0$ contained within the black hole.

Within our coordinate system, the Tolman temperature gradient drives $T (r) \rightarrow \infty$ as $r \rightarrow r_s$. The gradient is physical, but the divergence is an artifact of our fixed-background approximation; we have not allowed the spacetime to backreact. To produce Fig.~\ref{fig:A0}, we followed the example of Refs.~\cite{ThorneMembraneMNRAS,ThorneMembraneBook,MacDonaldMembrane} and evaluated the field with a boundary condition at a ``stretched horizon," $r_s + \epsilon$, rather than at $r_s$, with $\epsilon / r_s = 10^{-6}$. In forthcoming work \cite{EAMprep}, we address this limitation by solving the coupled Einstein field equations and equation of motion for $A_0^a (x)$ within the quasi-Abelian regime.

\section{Discussion}

We have generalized the description of Debye screening of charges in hot plasmas to curved spacetime backgrounds. For fluids undergoing normal flow in approximately static spacetimes, we found that the characteristic screening length $\lambda_D(x)=1/m_D(x)$ is set by a Debye mass $m_D(x)\sim g_s T(x)$ given in Eq.~(\ref{mDfinal}), where $g_s$ is the dimensionless gauge coupling. This Debye mass is the natural generalization of the Minkowski result, upon allowing the temperature to have spatial gradients due to gravitational redshift, thereby reproducing Tolman's classic result \cite{Tolman:1930ona,Tolman:1930zza,Santiago:2018lcy}.

To characterize Debye screening, we constructed an effective theory for long-wavelength excitations in a hot Yang-Mills plasma. We analyzed the dynamics of high-momentum particles in the plasma (with momentum $k_{\rm hard}\sim T$) by considering classical transport equations, and exploited the structure of Hamiltonian mechanics to show that non-Abelian self-interactions induce an effective local mass $m_D(x)$ for the $A_0^a$ components of the soft modes (with momentum $k_{\rm soft}\sim g_sT$).

We applied our results to solve for the gauge potential $A_\mu^a$ in a few cases of interest. We recovered the well-known Wu-Yang monopole solution in the Minkowski limit and generalized it to FLRW spacetimes, demonstrating the self-consistency of the quasi-static approximation in regimes of interest, for which cosmological time-scales may be neglected compared to $1/k_{\rm soft}$. 

Lastly, we analyzed Debye screening in the vicinity of a primordial black hole immersed in a hot quark-gluon plasma. Such black holes can form by absorbing regions of the plasma with nonvanishing net color charge, with characteristic size set by $\lambda_D$; hence the resulting primordial black holes can have a residual, net color charge. Building on the examples involving homogeneous spacetimes, we identified a regime in which the soft modes of the gauge field outside the black hole exhibit quasi-Abelian behavior, $A_\mu^a (x) \approx A_0^a (r) \delta_\mu^0$. Given the Tolman temperature gradients, the interaction strength $\alpha_s$ runs to smaller values near the event horizon. Incorporating this unusual example of asymptotic freedom, we solved numerically for the soft-mode gauge potential $A_0^a (r)$, and found enhanced screening of the charge enclosed within the black hole, due to an increased density of the plasma near the event horizon.

Our examples have been restricted so far to fixed background spacetimes. Future work will focus on exploiting our EFT to study realistic cosmological scenarios involving primordial black holes. This will require solving the coupled Einstein-Yang-Mills equations for a black hole in a hot, non-Abelian plasma \cite{EAMprep}. This formalism can also be used to consider primordial black holes with substantial QCD color charge, which could have formed at very early times in cosmic history, well before the QCD confinement transition \cite{Alonso-Monsalve:2023brx}.

\section*{Acknowledgements}

It is a pleasure to thank Chris Akers, Peter Arnold, Thomas Baumgarte, Jolyon Bloomfield, Daniel Harlow, Scott Hughes, Edmond Iancu, Mikhail Ivanov, Patrick Jefferson, Jamie Karthein, Hong Liu, Cristina Manuel, J\'{e}r\^{o}me Martin, Govert Nijs, Krishna Rajagopal, Phiala Shanahan, Vincent Vennin, and Xiaojun Yao for helpful discussions. Portions of this work were conducted in MIT's Center for Theoretical Physics and supported in part by the U.~S.~Department of Energy under Contract No.~DE-SC0012567. EAM is also supported by a fellowship from the MIT Department of Physics.


%

\end{document}